\documentclass[a4paper,11pt]{article}
\pdfoutput=1 

\usepackage{jcappub} 

\usepackage[T1]{fontenc} 
\usepackage[utf8]{inputenc}
\usepackage{hyperref}
\usepackage{xurl}   %

\title{\boldmath Ground Level Enhancement (GLE$\#$77) in the gamma-ray component: First observation from Arctic and Antarctic stations}

\author[a,1]{Pranali Thakur,\note{Corresponding author.}}
\author[a]{Geeta Vichare,}
\author[b]{Selvaraj Chelliah}

\affiliation[a]{Indian Institute of Geomagnetism (IIG),\\Sector-18, Navi Mumbai, India}
\affiliation[b]{Equatorial Geophysical Research laboratory, IIG,\\Tirunelveli, India}

\emailAdd{pranalispace@gmail.com}
\emailAdd{vicharegeeta@gmail.com}
\emailAdd{selvaegrl@gmail.com}

\abstract{This article presents the observations of the extreme ground-level enhancement (GLE $\#$77) of Solar Cycle 25 that occurred on 11 November 2025, using ground-based NaI(Tl) gamma-ray detectors deployed at Arctic and Antarctic stations, together with neutron monitor data and particle measurements from the GOES-18 satellite. The event was associated with an intense X-class solar flare and a strong solar energetic proton event. This paper reports the first ground-based detection of a GLE using gamma-ray detectors operating simultaneously in both polar regions, which are concurrent with increases in neutron monitor counts. Thus highlights the capability of polar gamma-ray detectors to complement traditional neutron monitor observations during extreme solar proton events. A detailed analysis revealed distinct prompt and delayed responses during the event evolution. Interestingly, the signature of the prompt peak of GLE$\#$77 (at 10:38 UT) was observed up to high-rigidity neutron monitors (low latitudes). However, the delayed peak (at 13:08 UT) was not seen at the stations with rigidity $>$ 6 GV. The timing of the prompt and delayed peaks coincided with the proton flux peaks observed by the GOES-18 satellite at energies $>$ 150 MeV and 12-99 MeV, respectively. It is observed that the GLE amplitude has a strong dependence on geomagnetic cut off rigidity and has a weak solar zenith angle dependence.}

\begin{document}
\maketitle
\flushbottom

\section{Introduction}
\label{sec:intro}
 Ground-level enhancements (GLEs) are rare and extreme manifestations of solar proton events (SPE) in which relativistic particles, primarily protons, penetrate the Earth’s atmosphere and generate secondary radiation detectable at ground level. These events are typically associated with major solar eruptions, such as X-class solar flares and fast coronal mass ejections (CMEs), and represent the highest-energy tail of the SPE spectrum. Forbush 1946 \citep{forbush1946three} first observed the high-energy particles from the sun as sudden increases in intensity in ground-level ion chambers during February and March 1942, and named these increases as ‘ground level enhancements’ (GLEs). Traditionally, GLEs have been identified and studied using the global neutron monitor (NM) network, which records sudden and statistically significant increases in secondary neutron count rates during SPE events. A formal definition of a GLE requires near-simultaneous enhancements at multiple neutron monitor stations, including at least one near sea level, together with a corresponding increase in space-borne proton flux measurements \citep{poluianov2017gle}. GLEs have provided a unique opportunity to investigate particle acceleration and transport processes operating at the Sun and in the heliosphere. To date, 77 GLEs have been documented (\url{https://gle.oulu.fi/#/} ; \url{https://www.nmdb.eu/nest/ }), spanning several solar cycles, with many events exhibiting complex temporal structures such as anisotropic onsets, delayed maxima, and multiple intensity peaks \citep{belov2005ground,mccracken2008investigation,moraal2012time,gopalswamy2012properties} and references therein. Studies have shown that some GLEs exhibit complex temporal structures, which are commonly interpreted in terms of different particle acceleration and transport processes. Early, highly anisotropic onsets are often associated with rapid acceleration near the flare site or in the lower corona, whereas delayed enhancements are generally attributed to particle acceleration at CME-driven shocks and subsequent propagation effects in the heliospheric magnetic field \citep{shea2012space,gopalswamy2012properties}.
Recently, Volkov et al.(2025) \citep{volkov2025observation} reported the observation of GLE $\#$77 using neutron detector arrays at the Experimental Complex NEVOD in Moscow, demonstrating good agreement with conventional neutron monitor observations. Furthermore, Chilingarian et al.(2025) \citep{chilingarian2025solar} presented detailed ground-based measurements of atmospheric neutrons and muons associated with GLE$\#$77 using the Aragats Solar Neutron Telescope. Their study revealed a clear dual-peak structure with an initial hard component at 10:28 UT and a softer component at 10:45 UT. They reported the first simultaneous energy spectra of secondary particles in the 40–600 MeV range during a GLE, confirming that the Sun was capable of accelerating protons to GeV energies during Solar Cycle 25. Despite these advances, ground-based studies of GLEs have so far been dominated by neutron and muon measurements, while gamma-ray observations at ground level remain largely unexplored. Gamma rays in the MeV energy range are sensitive to atmospheric cascades initiated by high-energy particles and can therefore provide complementary information on particle interactions, atmospheric modulation, and local geomagnetic effects. High-latitude regions are particularly favourable for such observations due to their low geomagnetic cutoff rigidity and reduced magnetic shielding.
In this paper, we investigate the extreme GLE$\#$77 that occurred on 11 November 2025 using ground-based NaI(Tl) gamma-ray detectors deployed at Arctic and Antarctic stations, together with neutron monitor observations and X-ray and proton flux measurements from the GOES-18 satellite. We focus on the temporal evolution of the gamma-ray flux during the event and examine neutron monitor responses. Special attention is given to the roles of geomagnetic rigidity and solar zenith angle, which may influence the observed gamma-ray and neutron signatures at different locations. This work presents the first ground-based gamma-ray detection of a GLE observed simultaneously in both polar regions.

\section{Data}
\label{sec:Data}

Recently, the Indian Institute of Geomagnetism (IIG), Mumbai, installed NaI(Tl) scintillation detectors at the Indian polar stations: Ny-Ålesund in the Arctic (Geographic coodinates: 76.63$^o$N, 11.87$^o$E; Rigidity: 0.01, Altitude: 40 m) and Maitri in Antarctica (Geographic coodinates: -67.85$^o$S, 11.43$^o$E; Rigidity: 0.3, Altitude: 117 m). Each NaI(Tl) detector is a rectangular cuboid with dimensions  10.16 cm $\times$ 10.16 cm $\times$ 40.46 cm . These NaI(Tl) detectors measure gamma-ray counts that include terrestrial as well as secondary cosmic rays produced in a cascade of extraterrestrial origin \citep{vichare2018equatorial,datar2020causes,datar2021barometric,thakur2024effect,thakur2025some}. The energies of the background radioactive nuclide due to terrestrial radioactivity are in the range of 300 keV to 3 MeV, and hence, the gamma ray with energy $>$ 3 MeV can be considered to originate from extra-terrestrial sources.  

The $X$-ray and particle flux data were obtained from the Geostationary Operational Environmental Satellite GOES-18 spacecraft 
(\url{https://data.ngdc.noaa.gov/platforms/solar-space-observing satellites/goes/goes18/l2/data/sgps-l2-avg1m/}). The spacecraft carries two Solar and Galactic Proton Sensor (SGPS) units mounted in the $-X$ and $+X$ directions. The SGPS$-X$ unit faces west, while the SGPS+$X$ unit faces east. Each SGPS unit consists of three solid-state silicon detector telescopes, T1, T2, and T3, which measure differential proton fluxes in the energy ranges 1-25 MeV, 25-80 MeV, and $>500$ MeV, respectively \citep{kress2021observations}. In this study, we have used Level-2 data products with a 1-minute temporal resolution for both $X$-ray irradiance and proton flux measurements.

The Neutron flux measurements recorded by the global neutron monitor (NM) network during GLE$\#$77 on November 11, 2025, are used in this analysis. Here, we have taken neutron monitor data from the NMDB database with a 2-minute resolution and from the OULU database with a 1-minute resolution. We use atmospheric pressure–corrected neutron counts rates sampled at a 2-minute resolution, from which a five-point moving average is calculated. The neutron and gamma-ray fluxes at each station are then normalized relative to their pre-event baselines. The reference level, $N_{mean}$, is defined as the average neutron and gamma-ray flux measured prior to the onset of the disturbance. $N_{t}$ is the neutron and gamma-ray counts at time t for a given observatory. The resulting normalized percentage variation is computed as,

\begin{equation} \label{eq1}
N_{norm} = \frac{\ N(t)-\ N_{mean}}{\ N_{mean}}*{100}
\end{equation}

\section{Observation of Ground Level Enhancement (GLE\#77) on 11 November 2025}
\label{sec:sec3}


 \begin{figure}[ht!]
 \includegraphics[width=1 \textwidth]{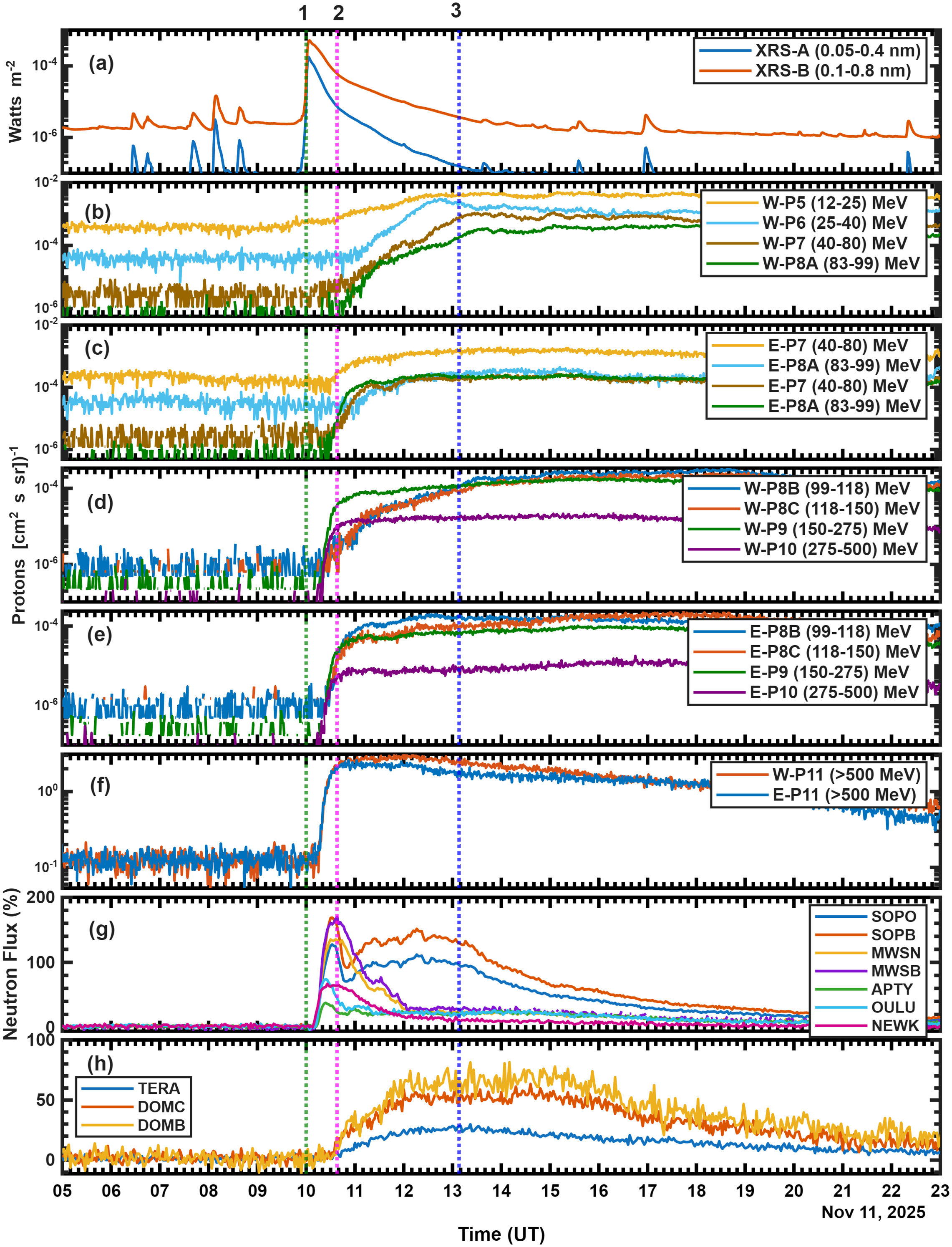}
\caption{GOES-18 solar flare X-ray fluxes, SGPS proton fluxes, and neutron fluxes during GLE~\#77 on 11 November 2025.}
\label{fig:Fig1}
\end{figure}

Active region AR 4274 produced an X5.1-class solar flare on 11 November 2025, with a start time at 09:49 UT, a peak at 10:04 UT, and an end time at 10:17 UT, as measured by the GOES-18 X-ray satellite. This intense X-class (X5.1) solar flare, which occurred on 11 November 2025, produced ground-level enhancement GLE$\#$77.

Figure \ref{fig:Fig1} presents the temporal evolution of the event using space-borne and ground-based observations on 11 November 2025. The X-ray and proton flux measurements from the GOES-18 satellite and neutron flux variations recorded by the global neutron monitor network (\url{https://www.nmdb.eu/nest/}) are shown. The top panel of Figure \ref{fig:Fig1} shows the soft X-ray irradiance measured by the GOES X-ray Sensor, clearly indicating the solar flare impulse at 10 UT (green dotted line marked as "1"). The plots in Figure \ref{fig:Fig1}(b) to \ref{fig:Fig1}(e) display the differential proton fluxes (P5–P10) recorded by the east and west-facing Solar and Galactic Proton Sensor units onboard GOES-18. Here, we have displayed the proton flux with energies $>$ 12 MeV, as lower energy flux (1-12 MeV) did not show any change during this event. Figure \ref{fig:Fig1} (f) shows the integral proton flux (P11, $>$500 MeV). The neutron flux measured at some NM stations is shown in Figure \ref{fig:Fig1}(g)-(h). A pronounced enhancement in proton flux above 12 MeV commencing at $\sim$10:10 UT is observed across multiple energy channels, up to several hundreds of MeV, indicating an arrival of high-energy solar energetic protons near Earth, due to the efficient acceleration of high-energy solar protons during the flare. It can be observed that the west-facing P5, P6, P7, and P8A show a gradual increase peaking around 13 UT Figure \ref{fig:Fig1}(b), whereas the same energy channels in the east-facing units peak at the earlier times Figure \ref{fig:Fig1}(c). Similarly, the variations in the W P8B and W P8C channels peak at the later times \ref{fig:Fig1}(d), while that in the east-facing channels is earlier \ref{fig:Fig1}(e). The higher energy protons ($>$ 150 MeV) show a rapid increase peaking near 10:38 UT (Figure \ref{fig:Fig1}(d) to (f)), while lower energy protons show a gradual increase peaking at 13:08 UT (Figure \ref{fig:Fig1}(b)).

\begin{table}
\centering
\caption{List of the Neutron Monitor Stations}
\label{tab:table1}
\begin{tabular}{cccccccc}
\hline
\hline

Sr & Station & Altitude & Rigidity  &  NM & Geographic & Geographic & Geomagnetic\\

 No. & Name  & asl & Rc & Type & Latitude &Longitude& Latitude \\
 
&  & (m) & (GV) &  &(deg) & (deg) & (deg) \\ 
\hline

1 & HLE1 & 3030 & 12.9 & 6NM64 & 20.72 N & 156.28 W & 21.49 N \\

2 & HLE2 & 3030 & 12.9 & 6NM64 & 20.72 N & 156.28 W & 21.49 N \\

3 & ICRB & 2373 & 11.5 & 3LND2061 bare & 28.3 N & -16.48 W & 33.10 N \\

4 & ICRO & 2373 & 11.5 & 3NM64 & 28.3 N & -16.48 W & 33.10 N \\

5 & ATHN & 260 & 8.53 & 6NM64 & 37.97 N & 23.78 N & 36.37 N \\
 
6 & MXCO &  2274 & 8.28 &  6NM64 & 19.33 N & 99.18 W & 27.53 N\\

7 & ROZH  &  1730	& 6.27 & LND2043BF3 & 41.43 N & 24.44 E&  39.66 N\\

8 & ROME  &  1730	& 6.27 & 20NM64 & 41.86 N & 12.47 E&  41.96 N\\

9 & AATB &  3340 & 5.90	 & 18NM64 &	43.14 N	& 76.60 E & 35.08 N \\

10 & BKSN &  1700 & 5.70 &	 6NM64 	& 43.28 N	& 42.69 E	& 38.82 N \\

11 & JUNG  & 3570 & 4.49& 18IGY   & 46.55 N	& 7.98 E & 47.25 N \\

12 & JUNG1	&  3475	& 4.49 & 3NM64  &	46.55 N	& 7.98 E & 47.25 N 	\\

13 & LMKS & 	2634 &	3.84 &	 8SNM64   &	49.20 N	& 20.22 E & 47.89 N	 \\

14 & IRKT &  435	& 3.64	&	 18NM64    &	52.47 N	& 104.03 E & 43.31 N	\\

15 & IRK2 &  2000	& 3.64	&	 12NM64    &	52.37 N	& 100.55 E & 43.25 N	\\

16 & IRK3 &  3000	& 3.64	&	 6NM64    &	51.29 N	& 100.55 E & 42.17 N	\\

17 & DRBS &  225	& 3.18	&	 9NM64    &	50.1 N	& 4.6 E & 51.24 N	\\

18 & NVBK &  163	& 2.91	&	 24NM64    &	54.48 N	& 83 E & 45.98 N	\\

19 & MOSC &  200	& 2.43	&	 24NM64    &	55.47 N	& 37.32 E & 51.48 N	\\

20 & NEWK &	50	& 2.40 & 9NM64   & 39.68 N	& 75.75 W  & 48.84 N  \\

21 &	KIEL2 & 	54 &	2.36 &	 18NM64 &	54.34 N &	10.12 E & 54.45 N	 \\

22 &	YKTK & 105 &	1.65	  & 18NM64  &	62.01 N	& 129.43 E & 53.39 N \\

23 &	KERG & 33 &	1.14	  & 18NM64  &	49.35 S	& 70.25 E & 56.3 S \\

24 &	CALG	& 1123	& 1.08 & 12NM64 	& 51.08 N	& 114.13 W &  57.48N 	\\

25 & OULU	& 	15 & 0.81	 & 9NM64  & 65.05 N	& 25.47 E & 62.32 N	\\

26 & SNAE	& 	856 & 0.73	 & 6NM64  & 71.4 S	& 2.51 W & 66.74 S	\\

27 & APTY &  181 &	0.65 & 18NM64   & 67.57 N &	33.4 E & 63.61 N	\\

28 & NRLK	&	0	& 0.63 & 18NM64  & 69.26 N &	88.05 E & 60.48 N	  \\

29 & TXBY	&	0	& 0.48 & 18NM64  & 71.36 N &	128.54 E & 62.66 N	  \\

30 & THUL & 26	& 0.30 & 9NM64   &	76.5 N &	68.7 W	& 85.55 N \\

31 & INVK &	21	& 0.30  & 6NM64  &	68.36 N &	133.72 W & 71.22 N  \\

32 & JBGO & 29 & 0.30	& 6NM64  &	74.6 S	& 164.2 E & 77.12 S	\\

33 & FSMT & 180 &	0.30 & 3NM64 & 60.02 N	& 111.93 W & 66.46 N  \\

34 & PWNK &  53	& 0.30 & 3NM64 	& 54.98 N	& 85.44W & 46.36 N\\

35 & MWSB	& 30	& 0.22  & 18NM64   & 67.60 S	& 62.88 E & 72.96 S	\\

36 & MWSN	& 30	& 0.22  & 18NM64   & 67.60 S	& 62.88 E & 72.96 S	\\

37 & MRNY	& 30	& 0.03  & 12NM64   & 66.55 S	& 93.02 E & 75.24 S	\\

38 & TERA &  32	& 0.01	 & 9NM64   &	66.65 S	& 140 E	& 73.54 S \\

39 & DOMB	& 3233	& 0.01	& Mini NM bare   & 75.06 S &	123.20 E & 83.35 S \\

40 & DOMC	& 3233	& 0.01	& Mini NM   & 75.06 S &	123.20 E & 83.35 S \\

41 & SOPB & 2820	& 0.10 & 3HeNM64  &	90 S & NA & NA\\

42 & SOPO  &  2820	& 0.10	& 3HeNM64  & 90 S	& NA	& NA\\

\hline
\end{tabular}
\end{table}


The increase in proton flux precedes the ground-level response and is followed by clear and significant enhancements in neutron monitor count rates at multiple stations worldwide. This confirms that relativistic solar protons penetrated the Earth’s atmosphere and produced secondary particles detectable at ground level. Accordingly, the event has been classified as GLE$\#$77 in the international ground-level enhancement catalogue maintained by the University of Oulu (\url{http://gle.oulu.fi/#/}). The combined satellite and ground-based measurements establish this event as one of the most energetic episodes of particle acceleration during Solar Cycle 25. The Global Neutron Monitor (NM) Network recorded notable increases in neutron flux with the onset between 10:10 and 10:20 UT on November 11, 2025, based on data from \url{https://www.nmdb.eu/nest/} and \url{http://gle.oulu.fi/#/}. The neutron fluxes measured at ground level by neutron monitors at different locations display both prompt and delayed peaks (Figure \ref{fig:Fig1}(g)-(h)). The prompt peak is short-lived, while the delayed peak has a longer duration compared to the prompt peak. The prompt enhancement observed in the neutron monitor matches with the sudden enhancement in higher-energy proton fluxes, as shown in the panels of Figure \ref{fig:Fig1} (e) and (f).  Hence, based on the neutron and proton fluxes, we have identified a prompt peak at 10:38 UT on 11 November 2026 (magenta dotted line that is marked as "2") and a delayed peak at 13:08 UT (blue dotted line marked as "3").

\subsection{Gamma-ray observations at Arctic}
\label{sec:sec3.1}
\begin{figure}[tbp]
 \centering 
 \includegraphics[width=0.6 \textwidth]{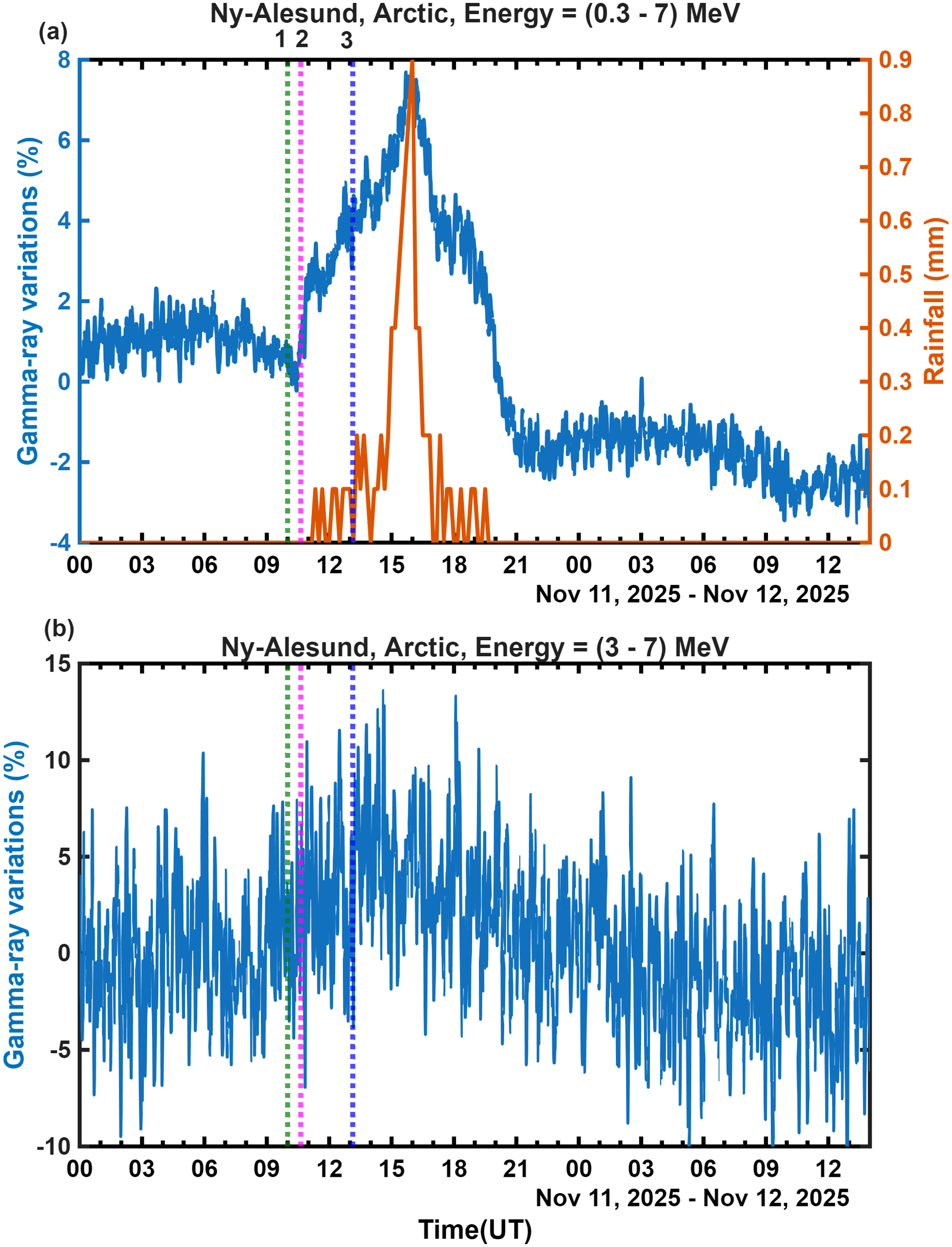}
\caption{Gamma-ray variations (300 keV to 7 MeV) and during 10-12 November 2025 (top panel); gamma-ray variation with energy 3-7 MeV during the same period (bottom panel)}
\label{fig:Fig2}
\end{figure}
The presence of these distinct features may suggest differences in particle transport, atmospheric cascade development, and viewing geometry at individual stations. Such variations reflect the combined effects of geomagnetic cutoff rigidity, local time, and propagation conditions in the heliospheric magnetic field. However, looking at the proton energy spectrum recorded by GOES-18, it appears that the energies might be responsible for prompt and delayed peaks.

Figure \ref{fig:Fig2}(a) shows the normalized percentage variation in gamma-ray data collected at the Ny-Ålesund station on 11 November 2025. The vertical lines are the same as those shown in Figure \ref{fig:Fig1}. Here, we present the gamma-ray flux in the entire energy range of 0.3–7 MeV. A sudden increase in the gamma ray at 10:38 UT (marked as "2") can be observed, which is similar to the enhancements in the neutron and proton flux discussed above. The enhancement at the second vertical line is approximately 1.5$\%$. The increase continued further beyond 13:08 UT (marked as "3") and peaked at around 16 UT. This is intriguing. There are several studies reporting rainfall-related enhancements in the low-energy gamma-ray counts \citep{barbosa2023precipitation,datar2020response}, which was attributed to the washout of terrestrial radionuclides during rainfall. These terrestrial gamma-rays have energies below 3 MeV \citep{datar2020response}. Therefore, we examined whether there was any rainfall on 11 November 2025. Interestingly, rainfall was recorded during this period. The rainfall/precipitation data recorded at Ny-Ålesund is shown by the dark orange curve in Figure \ref{fig:Fig2}(a). These measurements, with a 10-minute temporal resolution, were obtained from the Norwegian Centre for Climate Services (\url{https://seklima.met.no/observations/}). The rainfall started after 11 UT on 11 November 2025 and was maximum at 16 UT, thus coinciding with the time of maximum enhancement observed in the gamma-ray flux. Therefore, the precipitation at the site of gamma-ray observations might have contributed to the observed increase in gamma-ray counts at 16 UT, which is evident from Figure \ref{fig:Fig2}(a). Since the rain-related enhancement is confined to energies less than 3 MeV, we have examined the gamma-ray variations in the energy range above 3 MeV, which can be considered to be originating from the extra-terrestrial sources through secondary cosmic-ray cascades in the atmosphere. Figure \ref{fig:Fig2}(b) shows the $\%$ variation of the gamma rays in the 3-7 MeV energy range, and interestingly, the peak near 16 UT has disappeared. Though there are considerable fluctuations due to lower counts in the smaller energy range of 3-7 MeV, the amplitude of the gamma-ray variation within the vertical line marks is approximately 10$\%$.

\subsection{Gamma-ray observations at Antarctica}
\label{sec:sec3.2}

\begin{figure}[tbp]
 \centering 
 \includegraphics[width=0.7 \textwidth]{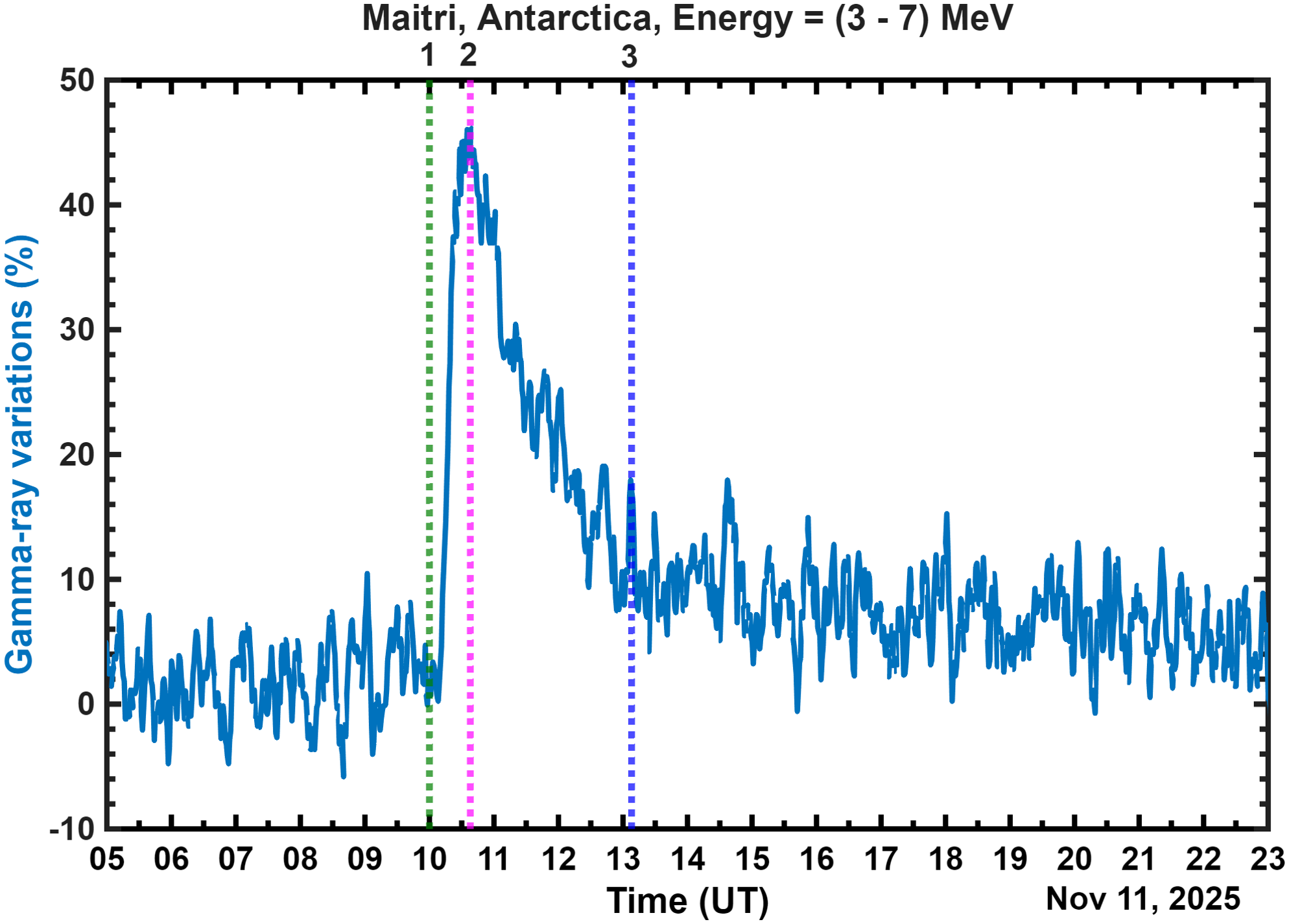}
\caption{Enhancement seen in gamma-ray flux with energy 3-7 MeV during 11 November 2025 at Maitri, Antarctica.}
\label{fig:Fig3}
\end{figure}

As discussed above in Section \ref{sec:sec3.1}, gamma rays in the energy range 3-7 MeV originate from extraterrestrial sources, such as GLE. Figure \ref{fig:Fig3} shows the normalized percentage variation of the gamma-ray data recorded at the Maitri station in Antarctica in the 3-7 MeV range. The vertical lines shown in the figure are the same as those used in Figure \ref{fig:Fig1} and \ref{fig:Fig2}. A clear increase associated with the GLE is seen in the gamma-ray observations recorded at Antarctica. The gamma-ray intensity increased by approximately 45$\%$ at 10:38 UT, related to the first peak of the GLE $\#$77, and later started decreasing without the occurrence of the delayed second peak. This signature matches well with the neutron flux variations at the MSWN, MSWB NM stations depicted in Figure \ref{fig:Fig1}(g), which are also located in Antarctica. To examine the possible influence of local weather, rainfall data from a nearby automatic weather station was examined, and no rainfall was recorded on 11 November.

\begin{figure}[tbp]
 \centering 
 \includegraphics[width=0.92 \textwidth]{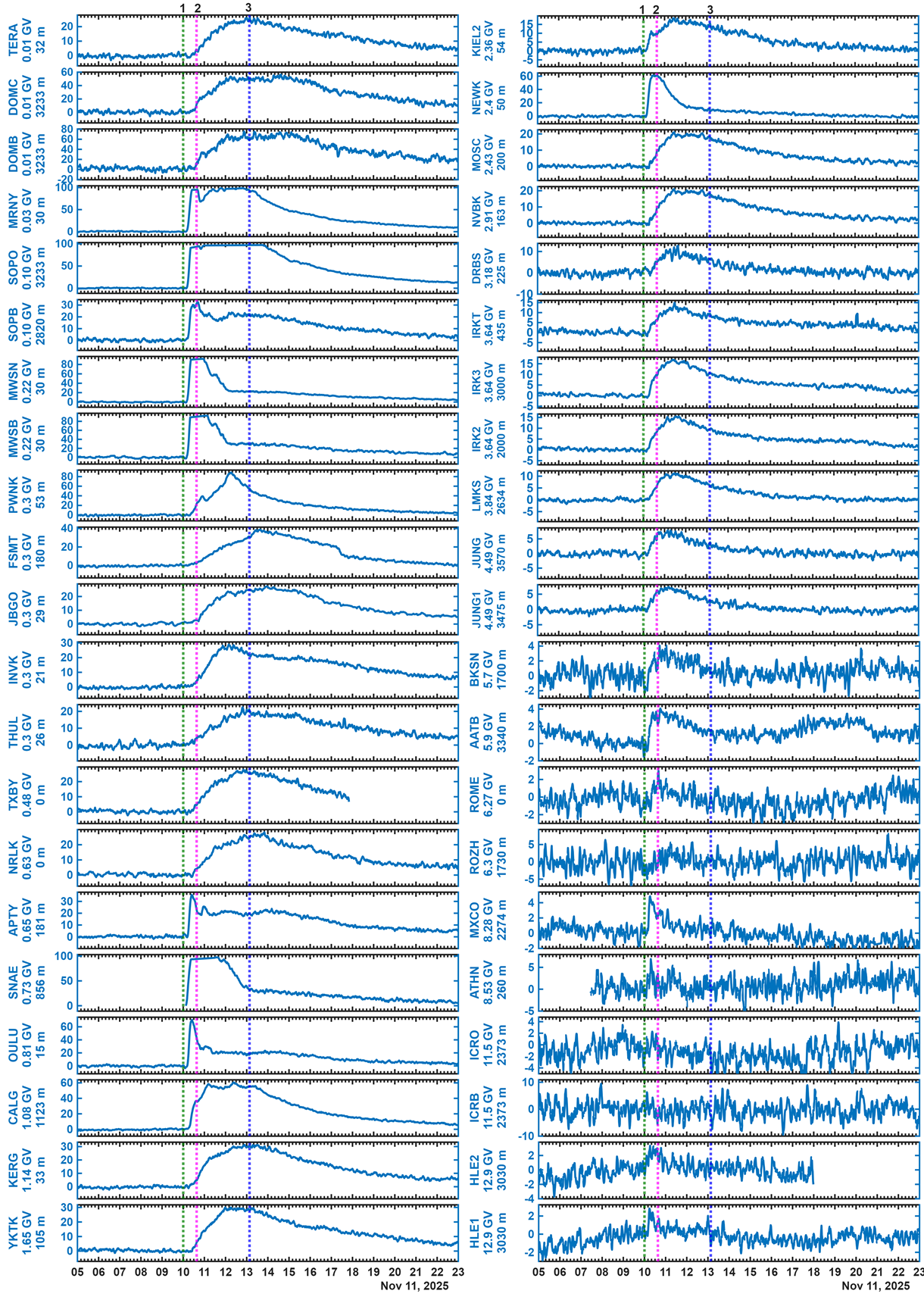}
\caption{Normalized neutron flux ($\%$) variations recorded at 42 neutron monitor stations, arranged from top to bottom in decreasing order of geomagnetic rigidity, during 11 November 2025 (GLE $\#$77).}
\label{fig:Fig4}
\end{figure}

\subsection{Neutron Monitor (NM) Observations}
\label{sec:sec3.3}

The NM data with a 1-minute time resolution, obtained from the OULU GLE database (\url{https://gle.oulu.fi/#/}), are shown in Figure \ref{fig:Fig4}. The list of NM stations used in this study, along with their geographic coordinates is provided in Table \ref{tab:table1}. The geomagnetic latitude of each station is calculated using the geomagnetic coordinate calculator available at \url{https://geomag.bgs.ac.uk/data_service/models_compass/coord_calc. html}. The geomagnetic cutoff rigidity values for the stations are taken from the NMDB website and are based on the compilation by \citep{gerontidou2021world}. 
Figure \ref{fig:Fig4} depicts the percentage variation in the neutron flux data for 42 worldwide NM stations during 11 November 2025. The five-point moving average of NM data of 1-minute time resolution is plotted in Figure \ref{fig:Fig4}, with increasing geomagnetic rigidity from top to bottom. Thus, the stations on the left-top have lowest rigidity and those on the right-bottom have the highest rigidity. Needless to mention that the vertical lines are the same as shown in Figure \ref{fig:Fig1}. It can be observed that the GLE signature varies at different stations. There are two types of GLE responses: One is prompt increase at 10:38 UT on 11 November 2025, and the other peak is gradual peaking at 13:08 UT which is also seen in Figure \ref{fig:Fig1}. Some stations are showing both the peaks, and some are showing only one dominant peak. Thus, the GLE signatures are complex at different stations. In general, GLEs are observed only at higher latitudes and are not been observed at the stations with higher rigidities (low latitudes). However, it can be observed from Figure \ref{fig:Fig4}, that the GLE$\#$77 related increase is clearly seen up to AATB or ROME station with geomagnetic rigidity of 5.9 GV and 6.27 GV respectively. Also, the NM stations like MXCO, ATHN show an increase during the event. This is normally not observed, and may be due to the intense nature of the present GLE, it is observed from the low latitudes as well. 

\section{Discussion and Results}
\label{sec:Discussion}

Solar energetic particles, typically reaching energies in the GeV range, can penetrate the Earth’s atmosphere and produce secondary particles detectable at ground level. The GLE event on 11 November 2025 represents the fourth and the strongest GLE of solar cycle 25 observed in the past 19 years. This GLE observed in neutron flux is comparable in intensity to GLE$\#$70 occurred on 13 December 2006. For the first time, the GLE is detected in the gamma-ray at polar regions and reported in this article. The gamma-ray responses observed at the Arctic and Antarctic stations during GLE$\#$77 are found to be different in terms of amplitude as well as the type of response. Although both stations recorded significant gamma-ray enhancements coincident with the GLE, the timing of the peak response differs between the two hemispheres. The Arctic station exhibits a predominantly delayed enhancement, whereas the Antarctic station shows a clear prompt peak. These features of gamma-ray observation at Antarctica and  Arctic of GLE are similar to the nearby NM stations of MWSB/MWSN and THUL, respectively. The geomagnetic cutoff rigidity at the Maitri station is approximately 0.3 GV, while that at Ny-Ålesund is about 0.01 GV. The amplitude of GLE signature in gamma-ray at Antarctica is higher (45$\%$) than that at Arctic (10$\%$). Also the amplitudes in the NM stations near arctic (THUL) is smaller ($\sim$20$\%$) than near Antarctica (MWSB/MWSN) is ($>$ 120 $\%$). It can be noted that the Antarctica was in the summer and Arctic was in the winter hemispheres. This motivated us to examine whether hemispheric differences, including geomagnetic rigidity, play a role in producing the observed prompt and delayed responses in the gamma-ray and neutron measurements. Therefore, the following sub-sections examine these dependences.

\subsection{Variation of amplitudes of prompt and delayed peaks with geomagnetic rigidity}

As shown in section \ref{sec:sec3.3}, the neutron monitor data show a complex response during this GLE event with either distinct prompt or delayed peak or both. In this sub-section, we plot the variation of the amplitudes of the prompt and delayed increases with the vertical geomagnetic cut off rigidity. Normally, GLEs are limited to the higher latitude stations and not observed at lower latitudes i.e. higher rigidity stations. However, due to strong intensity of the present GLE, it is observed even at lower latitudes with rigidities beyond 6 GV. This is something unique about the GLE$\#$77. As mentioned in the previous sections, the prompt and delayed peak are observed at 10:38 UT and 13:08 UT on 11 November 2025. We have identified the prompt peak at 10:38 UT and the delayed one between second and third vertical lines. That means the delayed peak may not be exactly at 13:08 UT, but any time after 10:38 UT. We have estimated the amplitudes of these peaks at different stations. It is cautioned that the amplitudes of the $\%$ increase might be influenced by the local factors such as the make of the detector, shielding, background counts, and local conditions, etc., and hence may not be the precise representation of the peak amplitudes. 

Figure \ref{fig:Fig5} shows the variation of the amplitudes of the prompt and delayed peaks with the geomagnetic rigidities. It can be observed that the amplitudes of these peaks have significant dependence on the rigidity. It decreases rapidly with increasing rigidity. This is obvious from the current understanding of the role of SPEs in GLE occurrences. The interesting thing to note from Figure \ref{fig:Fig5} is that at higher rigidity stations, only prompt peak is observed and the delayed peak is not seen there (zero amplitudes in Figure \ref{fig:Fig5}(b)). This may suggest that the prompt peak is due to the higher energy protons and delayed peak is due to relatively lower energy protons which do not have enough energies to penetrate through higher rigidities. From Figure \ref{fig:Fig1}, it is evident that the prompt peak coincides with the sudden increase in the proton flux of higher energies ($>$150 MeV), whereas the lower-energy proton flux reaches its maximum at delayed time at around 13:08 UT. This may indicate that the prompt peak in the neutron data is driven by high-energy protons which get sufficiently energize to reach the high rigidity locations, while the delayed peak likely includes a significant contribution from lower-energy protons.

Several studies have shown that some GLEs exhibit complex temporal structures, which are commonly interpreted in terms of different particle acceleration and transport processes. Early, highly anisotropic onsets were considered to be associated with rapid acceleration near the flare site or in the low corona, whereas delayed enhancements were generally attributed to particle acceleration at CME-driven shocks and subsequent propagation effects in the heliospheric magnetic field \citep{shea2012space,gopalswamy2012properties}. The relative contributions of these processes, however, remain an active area of research. Moraal $\&$ McCracken et al. 2012 \citep{moraal2012time} investigated all the GLEs of solar cycle 23, from GLE 55 on 6 November 1997 to GLE 70 on 13 December 2006, to study their morphology and pulse structure, and concluded that the associations between GLE magnitude with flare position, and CME dynamics do not yield a straightforward answer regarding the origin of the two pulses. Therefore, the present proposition of role of the protons with various energy levels in the prompt and delayed peaks may be a potential candidate for justifying the observed two peaks during GLE events.

\subsection{Variation of amplitude with solar zenith angle}

 \begin{figure}[tbp]
 \centering 
 \includegraphics[width=1 \textwidth]{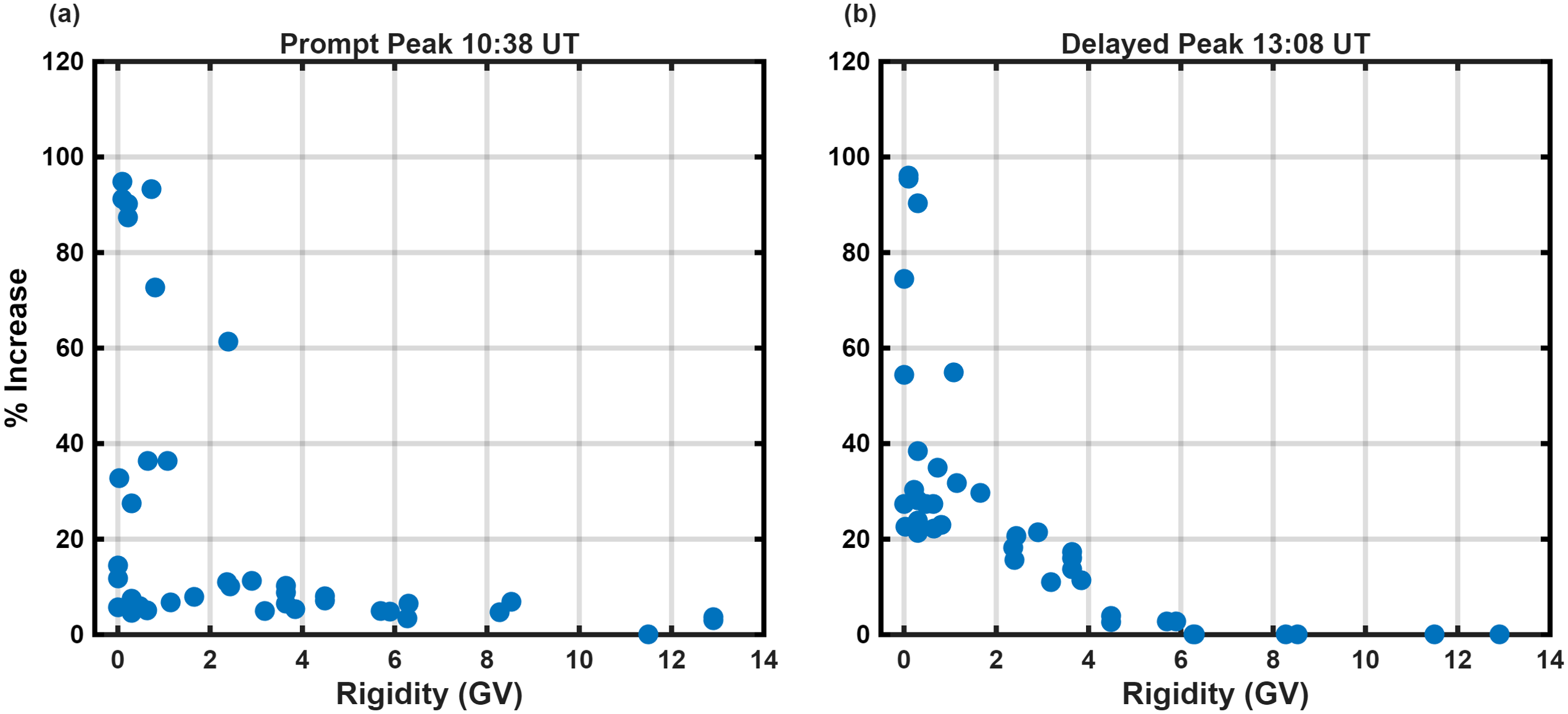}
\caption{The amplitude of $\%$ increase during (a) Prompt peak at 10:38 UT, (b) Delayed peak at 13:08 UT, during GLE$\#$77 on 11 May 2026.}
\label{fig:Fig5}
\end{figure}

 \begin{figure}[tbp]
 \centering 
 \includegraphics[width=1 \textwidth]{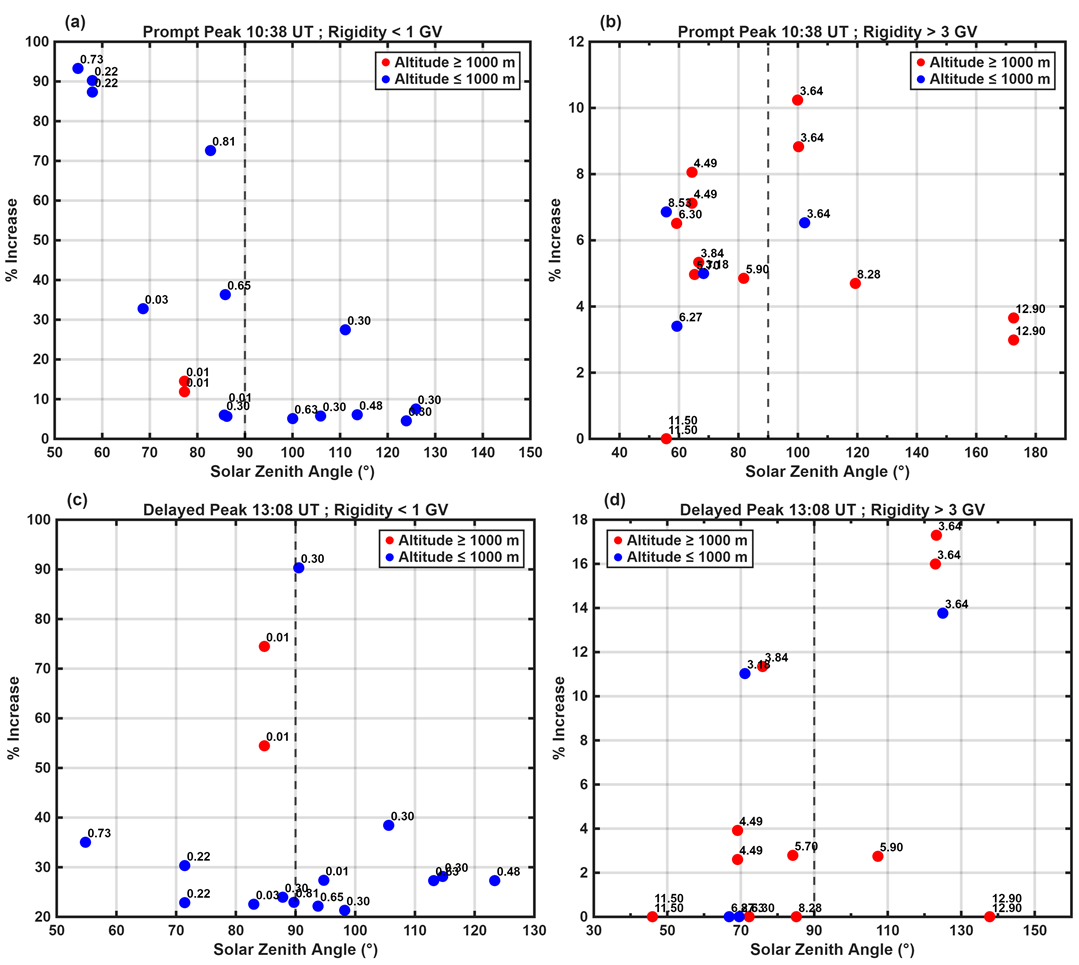}
\caption{Variation of amplitude of NM stations (a) Rigidity $<$ 3GV, (b) Rigidity $>$ 3 GV, with solar zenith angle of respective station during prompt peak at 10:38 UT; (c) Rigidity $<$ 3GV, (d) Rigidity $>$ 3 GV, with solar zenith angle of respective station during delayed peak at 13:08 UT. Red and blue points indicate the station located above and below 1000 m elevation.  The vertical black dashed line represents the solar zenith angle at 90$^o$.}
\label{fig:Fig6}
\end{figure}

From Figure \ref{fig:Fig5}, it is also evident that at lower rigidities, the scatter is very large, suggesting that in addition to the rigidity some other factor is also controlling the magnitudes of the GLE increase. The observed hemispheric asymmetry in gamma-ray behaviour at Arctic and Antarctica during GLE$\#$77 may suggest the sensitivity of ground-based gamma-ray detectors to local atmospheric conditions. At the time of the GLE onset, the Antarctic station was in the daylit hemisphere, characterized by a smaller solar zenith angle compared to the Arctic station. This may indicate that the factors such as solar zenith angle could play a role in modulating the amplitudes of GLE. Figure \ref{fig:Fig6} shows the variation of the amplitudes of prompt and delayed peaks with solar zenith angles. The solar zenith angles (SZA) are computed using \url{https://polaris.nipr.ac.jp/~aurora/cgi-bin/wdcc2/calSZA.cgi}. The scatter plots in Figures \ref{fig:Fig6}(a) and \ref{fig:Fig6}(b) are for the prompt peak using stations with rigidity $<$ 1 GV and $>$ 3 GV respectively. These boundaries are decided from Figure \ref{fig:Fig5}(a), as it appears less dependence on stations of rigidity $>$ 3GV, and lot of scatter is seen $<$ 1GV. Since the SZA is measured from the vertical axis, we marked an angle of 90$^o$ (a vertical dashed line) to indicate the boundary between the day and night regions. Similar scatter plots for the amplitudes of the delayed peaks are shown in Figure \ref{fig:Fig6}(c) and \ref{fig:Fig6}(d). Further, since the altitude of the station can be important in the intensity of the secondary cosmic ray particles \citep{thakur2025some}, we have indicated this information by red and blue coloured symbols. The symbol labels indicate the geomagnetic rigidity of a given station.
It can be observed from Figure \ref{fig:Fig6}(a) that in general, the stations with SZA$>$ 90$^o$ i.e. stations from the night region have lower amplitudes despite of lower rigidities. In Figure \ref{fig:Fig6}(b), this dependence is not that strong, but for the delayed peak there is no such dependence (Figure \ref{fig:Fig6}c and \ref{fig:Fig6}d). This suggests that the sunlit hemisphere is important for the prompt peak, not for the delayed peak. Also, it appears that the altitude of the station do not have clear role in the GLE amplitudes. Here, we would like to provide a caveat that the amplitudes are highly sensitive to the instrumental setup, including shielding, voltage, efficiency, etc., e.g., ICRO $\&$ ICRB are located at the same place, but one detector has lead shielding and the other one is lead-free. Therefore, the interpretations of the results presented here should be considered with caution.

\section{Conclusions}
\label{sec:summary&conclusion}
The Ground Level Enhancement (GLE $\#$ 77) on November 11, 2025, that was associated with an intense X-5.1 class solar flare and strong solar proton event (SPE), is the most outstanding example of a double peak structure in cosmic ray data. The key findings from this GLE event are summarized below:
\begin{itemize}
	\item A clear prompt peak at 10:38 UT and a delayed peak at 13:08 UT were observed in the worldwide network of NM stations and also in the gamma-ray detectors located at Northern and Southern polar latitudes.
    \item Signature of prompt peak of GLE was clearly observed in low latitude (high rigidity) neutron monitors.
    \item The delayed peak was not seen at the stations with rigidity $>$ 6 GV.
	\item Prompt and delayed peak timings coincided with the peaks in the proton flux with energies $>$ 150 MeV and 12-99 MeV, respectively, observed by GOES-18 satellite.
    \item It is found that the GLE amplitude has a strong dependence on the geomagnetic cutoff rigidity and a weak dependence on the solar zenith angle.
	\item This is the first GLE event recorded by a Gamma-ray detector simultaneously in the Arctic and Antarctic.

\end{itemize}

\acknowledgments

We acknowledge the NMDB database (\url{https://www.nmdb.eu/nest/}), established under the European Union’s FP7 program (contract no. 213007) and OULU GLE database (\url{https://gle.oulu.fi/#/}) for providing neutron monitor data. We also acknowledge the GOES-R program for access to the SGPS data set. The NaI(Tl) experimental setup at Ny-Ålesund, Arctic and Maitri, Antarctica, are operated by the Indian Institute of Geomagnetism (IIG). This work is supported by the Department of Science and Technology, Government of India.






\bibliographystyle{unsrt}
\bibliography{ref1}



\end{document}